# Collective Blinking of Upconversion Emission in Lanthanide-doped Nanocrystals

Tianzi Ma[1,3], Dingxin Huang[2,3], Yongjun Meng[1], Jianwei Tang[1*], Feng Li[2], Yihao Yu[1], Xiaorong Zhang[2], Guanying Chen[2*], Xue-Wen Chen[1*]

[1]School of Physics, Institute for Quantum Science and Engineering and Hubei Key Laboratory of Gravitation and Quantum Physics, Huazhong University of Science and Technology, Luoyu Road 1037, Wuhan 430074, China

[2]MIIT Key Laboratory of Critical Materials Technology for New Energy Conversion and Storage, School of Chemistry and Chemical Engineering & Key Laboratory of Micro-systems and Micro-structures, Ministry of Education, Harbin Institute of Technology, 150001 Harbin, China.

[3]These authors contributed equally

*To whom correspondence should be addressed:
J. T. (jianwei_tang@hust.edu.cn),
G.C. (chenguanying@hit.edu.cn ),
X.-W. C. (xuewen_chen@hust.edu.cn).

## Abstract

Fluorescence blinking, often regarded as a limitation for stable emitters, can enable super-resolution localization microscopy and serve as a versatile reporter of the photophysical states of quantum emitters and their interactions with local environment. However, conventional blinking emitters are typically single quantum systems with Stokes-shifted fluorescence, making them susceptible to autofluorescence background, weak signal, and irreversible photodegradation under prolonged excitation. In contrast, single lanthanide-doped upconversion nanocrystals are effectively background-free anti-Stokes emitters and demonstrate robust resistance to photodegradation, yet they are generally considered non-blinking owing to the presence of a large ensemble of uncorrelated emitting lanthanide ions within a single nanocrystal. Here we report the discovery and control of collective blinking in the upconversion luminescence of thousands of lanthanide ions within a single nanocrystal. The blinking exhibits on-off intensity ratio exceeding 10, persists for over 15 hours (over 10,000 cycles) without discernible photodegradation, and can be reversibly controlled by adjusting the excitation power. We elucidate a universal, activator-independent upconversion blinking mechanism, whereby a single quencher, stochastically generated via a cooperative multi-ion process, can intercept delocalized excitation energy within the $Yb^{3+}$ sensitizer network and darken the whole nanocrystal. Benefiting from the high-contrast, long-term photostable blinking and background-free upconversion emission, we achieve low-power super-resolution localization microscopy that resolves individual nanocrystals in dense aggregates with 1.2 nm precision. This work establishes a general strategy to realize and control collective blinking in photostable multi-emitter nanosystems, opening new opportunities in nanoscience, bioimaging, and quantum technologies.

## Introduction

Fluorescence intermittency was originally identified as a manifestation of quantum jumps in isolated single quantum systems, such as atoms, ions or molecules[1-4]. To date, blinking behavior has been observed across a wide range of natural and artificial single-emitter systems, including organic dyes[5-8], semiconductor quantum dots[9-13], fluorescent proteins[14,15], carbon nanodots[16,17], and color centers in solids[18]. The underlying mechanisms have since been found to extend well beyond simple quantum jumps, encompassing complex photophysical processes, involving charge trapping[12], defect dynamics[19], and environmental interactions[20,21]. Although blinking is often considered undesirable, analysis of blinking dynamics reveals the photophysical states of emitters and their interactions with local environment. For instance, at the molecular level, blinking dynamics serves as a sensitive reporter of local environmental fluctuations[20], protein conformational dynamics[22], and DNA hybridization events[23]. Blinking also underpins multiplexed detection[24], dynamic unclonable cryptographic primitives[25], and super-resolution localization microscopy[26-32], highlighting its versatility across nanoscience and biotechnology.

Normally, blinking cannot be observed in multi-emitter systems due to ensemble averaging, which averages out stochastic intensity fluctuations from individual emitters. Nevertheless, collective blinking can emerge in some specially designed multi-emitter architectures, such as conjugated polymers[33-35], J-aggregates[36], and counterion-assembled dye-doped nanoparticles[37]. In these systems, a stochastically formed quenching site quenches the entire ensemble through efficient energy migration, analogous to the energy-transfer process in natural light-harvesting complexes[38]. Compared with single-emitter systems, multi-emitter particles offer higher brightness and lower excitation thresholds. Yet they remain limited by autofluorescence background as Stokes-shifted fluorescence systems, and are prone to photodegradation due to cumulative photobleaching of the constituent emitters. Lanthanide-doped upconversion nanocrystals (UCNCs) represent a unique class of multi-emitter systems that efficiently convert low-energy photons into high-energy photons via multiphoton processes[39]. Such anti-Stokes emission allows complete rejection of the autofluorescence background, making UCNCs as background-free emitters[40]. Importantly, the inorganic dielectric host lattice and the shielding of 4f electrons endow UCNCs with outstanding resistance to photodegradation[41,42]. However,

UCNCs are generally considered non-blinking[43] owing to ensemble averaging among thousands of uncorrelated emitting lanthanide ions within a single nanocrystal.

Here we report the discovery and control of collective blinking in the upconversion luminescence (UCL) of lanthanide-doped nanocrystals. Our UCNC sample architecture features an ultrathin shell of activator ions ($Tm^{3+}$) within a continuous sensitizer-ion ($Yb^{3+}$) network. In isolated single UCNCs, UCL blinking exhibits on-off intensity ratios exceeding 10 and persists for over 15 hours without noticeable photodegradation, with dynamics that are reversibly tunable by excitation power. Mechanistic analysis across $Tm^{3+}$-, $Er^{3+}$-, and $Ho^{3+}$-doped systems shows that the blinking originates from activator-independent quenchers generated through high-order $Yb^{3+}$-$Yb^{3+}$ cooperative upconversion and acting by depleting excitation energy delocalized within the UCNC through energy migration among the $Yb^{3+}$ ions. By leveraging the favorable blinking characteristics of these UCNCs, including high on-off blinking contrast, long-term photostability, and low background, we demonstrate the super-resolution imaging that robustly resolves and localizes individual nanocrystals in dense aggregates with ~1.2 nm precision.

**UCL blinking with designed UCNCs**

The UCNCs studied here are typical sensitizer-activator co-doping systems[39], where ytterbium ions ($Yb^{3+}$) serve as sensitizers while thulium ions ($Tm^{3+}$) function as activators. The sensitizers absorb incident near-infrared photons and subsequently transfer the energy to the activators, populating higher-lying states that enable anti-Stokes emission. As shown in Fig. 1a, our UCNC design features an optimized core@shell@shell@shell architecture[44-46] ($NaYbF_4$ @ $NaTmF_4$ @ $NaYbF_4$ @ $NaYF_4$) with an overall diameter of approximately 30 nm. The core (5 nm in radius) and the second-outer layer (6.5 nm thick) are both doped with 100 mol% $Yb^{3+}$. In between, an ultrathin shell of $NaTmF_4$ (100 mol% $Tm^{3+}$, 0.5 nm thick) sever as the activator layer. The outermost shell (3 nm thick) is undoped $NaYF_4$, which acts as a passivation layer to isolate the inner layers from surface defects and environmental disturbances[47,48]. The details for synthesis and characterization of the UCNC samples can be found in Supplementary Note 1 and Extended Data Fig. 1.

For single-particle experiments, UCNCs are mono-dispersed on a glass coverslip via spin-casting, and optically characterized using a homebuilt microscope (see Supplementary Note 2 and Extended Data Fig. 2 for the detailed experimental setup). A continuous-wave laser beam at the wavelength of 980 nm is focused through a high-numerical-aperture objective to excite the samples. Subsequently, the UCLs are collected by the same objective, spectrally separated into the long-wavelength (~800 nm) and short-wavelength (~450 nm) emission bands, which are recorded simultaneously using two single-photon avalanche photodiodes. As displayed in Fig. 1b, the UCL time traces exhibit dramatic binary blinking behavior with an on-off intensity ratio of 13. Notably, the observed blinking events in the long- and short-wavelength UCL bands are temporally synchronous and have a consistent on-off intensity ratio, suggesting that all emission channels of $Tm^{3+}$ are affected simultaneously and thus the blinking originates from a process independent of the photophysical status of $Tm^{3+}$.

To establish the reproducibility of the observed UCL blinking, we have examined 121 additional UCNCs, all showing similar blinking behavior. Figure 1c presents the statistical distribution of the measured on-off intensity ratios for the full set, yielding a mean value of 13.6, with some exceeding 25, remarkably high even compared with single-emitter blinking systems. Supplementary Movie 1 provides a direct wide-field visualization of the UCL blinking of numerous isolated UCNCs over a continuous 15-hour duration. Notably, the blinking behavior remains robust under such prolonged continuous excitation. As an example, Extended Data Fig. 3 plots the UCL time trace of a single UCNC for over 15 hours, which includes over 10,000 blinking cycles and does not show any discernible degradation of the on-state intensity. Quantitative evaluation of the pixel intensities extracted from Supplementary Movie 1 reveals a signal-to-background ratio of 150 (Supplementary Note 3), demonstrating the effective suppression of the autofluorescence background.

**Mechanism of UCL blinking**

We propose that the observed UCL blinking arises from the stochastic formation and annihilation of quenchers. Although different blinking events may involve different quenchers, the observed single-step transitions indicate that each cycle is dominated by only one single quencher, which is fundamentally different from the photoswitching behavior recently reported in avalanching nanoparticles (ANPs)[49]. As schematically described in Fig. 1d, when the nanocrystal is in the normal emissive state (on-state), the sensitizers ($Yb^{3+}$) harvest pump photons and mediate the migration of the excitation energy through the continuous sensitizer network, followed by transfer to the activators ($Tm^{3+}$) to generate UCL. The transition from the on-state to the off-state is triggered by the generation of a quencher in the nanocrystal. The quencher acts as an energy sink, intercepting the delocalized excitations migrating through the sensitizer network and thereby suppressing the UCL of the nanocrystal. The UCNC returns to the on-state upon the spontaneous annihilation of the quencher. This "on-off" cycle repeats stochastically, generating the observed blinking behavior. The sub-nanometer thickness of the activator interlayer, combined with the high $Yb^{3+}$ concentration in the adjacent layers, preserves a continuous, high-density sensitizer network, enabling excitation energy to migrate throughout the nanocrystal and become spatially delocalized. By incorporating an inert $Y^{3+}$ passivation shell to isolate surface defects and an ultrathin $Tm^{3+}$ layer to limit the number of activators, the quenching rate introduced by a single quencher could surpass the combined rates of all other decay channels. Consequently, a single quencher can intercept the delocalized excitations and effectively darken the entire nanocrystal.

To examine the generality of our proposed blinking mechanism, we replace $Tm^{3+}$ with $Er^{3+}$ or $Ho^{3+}$ ions in the ultrathin activator layer of the UCNCs. Specifically, 5 mol% $Er^{3+}$-doped UCNCs and 5 mol% $Ho^{3+}$-doped UCNCs are synthesized with the identical core@shell@shell@shell architecture. A relatively low activator concentration of 5 mol% is adopted in both systems to ensure a pronounced on-off blinking contrast. As depicted in Fig. 2a and 2b, respectively, the UCL time traces of $Er^{3+}$- and $Ho^{3+}$-doped UCNCs indeed exhibit evident blinking, temporally synchronous across different UCL spectral bands.

Such universal UCL blinking behavior observed across different types of activators ($Tm^{3+}$, $Er^{3+}$, and $Ho^{3+}$) supports the generality of the proposed mechanism and directs attention to the $Yb^{3+}$ sensitizer network shared by all three systems. We therefore examine the Stokes-shifted or downconversion luminescence (DCL) from the sensitizers ($Yb^{3+}$) in a single UCNC under constant illumination (Supplementary Note 2.2). As plotted by the red traces in Fig. 2c and Fig. 2d, the measured DCLs of $Yb^{3+}$ in both $Er^{3+}$- and $Ho^{3+}$-doped UCNCs indeed exhibit temporal fluctuations synchronous with the recorded UCLs. In $Tm^{3+}$-doped UCNCs, the DCL from $Yb^{3+}$ turns out to be too weak at the single-particle level for meaningful analysis. The synchronous blinking of $Yb^{3+}$ DCL and activator UCL represents a direct evidence that the quencher intercepts the delocalized excitation energy within the $Yb^{3+}$ sensitizer network, rather than acting on the activators.

We further perform quantitative analysis on the DCL decay dynamics at on- and off-states (Supplementary Note 2.3). The color-coded traces in Fig. 2e and Fig. 2f present the DCL decay curves of the on-state (orange) and off-state (yellow) in a single $Er^{3+}$-doped and $Ho^{3+}$-doped UCNC, respectively. We extracted the decay rates of the excited $Yb^{3+}$ in the on- and off-states, denoted as $\Gamma_{\mathrm{on}}$ and $\Gamma_{\mathrm{off}}$, respectively. At the on-state, the decay rate is given by $\Gamma_{\mathrm{on}} = \Gamma_{\mathrm{s}} + \Gamma_{\mathrm{a}}$, where $\Gamma_{\mathrm{s}}$ and $\Gamma_{\mathrm{a}}$ represent the spontaneous decay rate and the energy transfer rate to activators, respectively. In the off-state, the presence of a quencher creates an additional nonradiative decay channel with a rate $\Gamma_{\mathrm{q}}$, leading to an accelerated total decay rate $\Gamma_{\mathrm{off}} = \Gamma_{\mathrm{s}} + \Gamma_{\mathrm{a}} + \Gamma_{\mathrm{q}}$. Notably, although the absolute values of $\Gamma_{\mathrm{on}}$ and $\Gamma_{\mathrm{off}}$ differ between $Er^{3+}$-doped and $Ho^{3+}$-doped systems due to their different $\Gamma_{\mathrm{a}}$, the extracted quenching rates ($\Gamma_{\mathrm{q}} = \Gamma_{\mathrm{off}} - \Gamma_{\mathrm{on}}$) of the two systems exhibit excellent consistency, amounting to $(198.2\ \mu s)^{-1}$ and $(195.8\ \mu s)^{-1}$ for $Er^{3+}$ and $Ho^{3+}$ system, respectively. We note that the variation of the DCL decay rates also shows consistency with the on-off intensity ratio. Specifically, the measured decay rate ratios $\Gamma_{\mathrm{off}}/\Gamma_{\mathrm{on}}$ (2.3 for the $Er^{3+}$ system; 4.5 for the $Ho^{3+}$ system) are in excellent agreement with the measured intensity ratios $I_{\mathrm{on}}/I_{\mathrm{off}}$ (2.3 for the $Er^{3+}$ system; 4.4 for the $Ho^{3+}$ system). The nearly identical quenching rates for the $Er^{3+}$- and $Ho^{3+}$-activated systems indicate a shared quencher species and quenching process that is independent of the activator identity.

We further probe the quencher-generation mechanism by investigating the influence of the excitation power density on the UCL blinking dynamics of Tm-doped UCNCs. At low excitation power densities, the UCNC displayed stable, nonblinking UCL as conventional UCNCs (Extended Data Fig. 4). The onset of UCL blinking is observed at an excitation threshold of about 1.9 $kW/cm^2$. Above this threshold, the blinking frequency increases and the fraction of time when the UCNC resides in the on-state decreases progressively with the increase of the excitation power density. Figure 3a shows the evolution of the UCNC blinking behavior during an excitation power cycle, in which the excitation power density is first gradually increased from 1.9 kW/cm$^2$ to 2.7 kW/cm$^2$ and then decreased back to 1.9 kW/cm$^2$ (in steps of 0.2 kW/cm$^2$). We have repeated the excitation-power modulation cycle shown in Fig. 3a for 20 times over a period of 3 hours (see Extended Data Fig. 5 for the intensity trace). Figure 3b plots the evolution of the on-state time fraction, which exhibits a cyclic response in accordance with the change of the excitation power density. From these cyclic measurements, we obtain the mean on-state time fraction as a function of the excitation power density as displayed by the right panel of Fig. 3b. Notably, a modest increase in the excitation (42%, from 1.9 to 2.7 kW/cm$^2$) led to a sharp rise in the off-state fraction from 8% to 61%. This highly nonlinear power dependence suggests that quencher generation is governed by a high-order multiphoton process, most likely via energy accumulation among multiple excited $Yb^{3+}$ ions, consistent with the well-documented $Yb^{3+}$-$Yb^{3+}$ cooperative upconversion process[50]. Such multi-ion cooperative process can induce electron ionization, and the released electron may be subsequently trapped by a defect within $NaYbF_4$ crystal to form an optically active quencher[51].

**Super-resolution microscopy with UCL blinking**

As a proof of concept, we next leverage UCL blinking for super-resolution localization microscopy. The high on-off blinking contrast, combined with the intrinsically low autofluorescence background afforded by anti-Stokes emission, substantially enhances the signal-to-noise ratio compared to conventional down-conversion blinking emitters. In addition, the exceptional photostability of the UCNCs allows extended time of data acquisition, which is essential to accumulate sufficient localization events for high-resolution imaging and long-term tracking. As shown in Fig. 4, we apply these advantages to resolve and localize individual nanocrystals in aggregates randomly formed during sample preparation.

Before presenting the results, we briefly introduce an optimized localization algorithm compatible with high-duty-cycle blinking dynamics (see Supplementary Note 4 and Extended Data Fig. 6 for details). We first record a time series of wide-field UCL image frames (Supplementary Movie 2) registered as a three-dimensional (3D) $(x, y, t)$ image matrix as shown in the middle panel of Fig. 4a. Four representative frames at different times are displayed on the left panel of Fig. 4a, where three spots (denoted as “I”, “II”, and “III”) are much brighter than the others, indicating the presence of multiple luminescent UCNCs within each spot. Subtracting each frame with the preceding one yields a time-differential image sequence represented by another 3D ($x$, $y$, $t$) matrix. In this 3D time-differential image matrix, clusters of positive (red disk) and negative (blue disk) matrix elements correspond to the “turn-on” and “turn-off” blinking events of individual UCNCs within a luminescence spot, respectively, as shown in Fig. 4b. Intervals between temporally adjacent blinking events represents stable emission intervals without blinking, as exemplified by the columnar regions $N_1$ and $N_2$ in Fig. 4b. $N_1$ and $N_2$ are also marked in the 3D image matrix in Fig. 4a. Averaging the image frames over $N_1$ and $N_2$ yields luminescence spots $S_1$ and $S_2$. Since $S_1$ and $S_2$ represent the images of spot “I” differ by one blinking event, their differential image ($S_1$ - $S_2$) isolates the luminescence spot of a single UCNC. Fitting this difference spot to a 2D Gaussian function yields the spatial coordinates of the UCNC with nanometer localization precision. Unlike previous blinking-based methods relying on a single-frame substruction[26,27,30], our algorithm identifies the stable emission intervals and accumulate their signal prior to differencing. By integrating photons within the extended intervals, it maximizes the photon budget for each localization, substantially increasing both the signal-to-noise ratio and the resulting localization reliability.

The localizations are corrected for sample drift and compiled into a 2D point cloud. By applying Gaussian rendering[52] on this point cloud, we reconstruct the super-resolution localization image. Figure 4c displays the optically resolved image of the UCNC sample, which agrees remarkably well with the corresponding AFM image shown in Fig. 4d. Almost every UCNC resolved in the AFM image is successfully identified in the optical localization image. Magnified views of the three densest regions (“I”, “II”, and “III”) are presented in the three corresponding panels of Fig. 4e. For each UCNC, there is a circle of obtained localization points, which are clustered and further used to determine the position of the UCNC. We obtain localization precisions of 1.2 nm in $x$ and $y$ directions (Supplementary Note 5, Extended Data Table 1). Our optical localizations excellently match the AFM results even within the dense UCNC aggregates. Notably, in region “I”, the optical approach successfully resolves and localizes eight distinct nanocrystals, even though these nanocrystals are barely decipherable in the corresponding AFM topography image.

## Conclusion

In summary, we have discovered and systematically characterized collective UCL blinking in lanthanide-doped UCNCs, overturning a long-standing perception that the UCNCs are non-blinking optical emitters. We have also elucidated the universal UCL blinking mechanism originating from stochastic creation and annihilation of optically active quenchers. By incorporating an inert $Y^{3+}$ passivation shell to isolate surface defects and an ultrathin activator layer to limit the number of activators, a single quencher can darken the whole nanocrystal. Synchronous blinking of $Yb^{3+}$ DCL and activator UCL serves as direct evidence that quenchers intercept delocalized excitation energy within the $Yb^{3+}$ sensitizer network rather than acting on the activators. A share quenching rate for samples with different activators indicates activator-independent quencher species and quenching process. The strong nonlinear power dependence of blinking additionally implies that quenchers are most likely generated through $Yb^{3+}$-$Yb^{3+}$ cooperative multiphoton ionization and subsequent charge trapping at crystal defects. Our blinking UCNC sample possesses exceptional performance, featuring high on-off intensity ratios up to 25 and extraordinary long-term photostability under constant illumination. By leveraging the UCL blinking possessing such properties and an event-differential imaging algorithm, we achieve robust super-resolution localization microscopy of individual UCNCs in dense aggregates with precisions of 1.2 nm. Notably, compared with previous UCNC-based super-resolution approaches[49,53-55], our method operates at a markedly lower excitation power density of ~ 2.3 kW/cm$^2$, making it particularly suitable for low-power imaging scenarios.

Our experiments have also demonstrated that the UCL blinking dynamics can also be reversibly controlled by adjusting the excitation intensity. We anticipate that the integration of these blinking UCNCs with optical nanocavities will exhibit richer photophysics and greater controllability[25,56,57]. The discovery and control of collective blinking in UCNCs unlocks the full potential of upconversion photonics for extensive applications relying on stochastic luminescence fluctuations, such as single-particle sensing[58], multiplexed detection[24], and dynamic optical cryptography[25]. Our work also provides a unique model system for investigating photophysics of energy migration, delocalization, and quenching in nanoscale solids, and establishes a new design paradigm for collective blinking in multi-emitter solid-state systems, stimulating transformative advances in nanoscience[59], bioimaging[60,61], and quantum technologies[62].

During the review process, we note two studies[63,64] related to this work.

**Data availability**

The data that support the plots within this paper and other finding of this study are available from the corresponding authors upon reasonable request.

**Code availability**

All codes used in this paper are available from the corresponding authors upon reasonable request.

**Acknowledgements**

We gratefully acknowledge financial support from the National Natural Science Foundation of China (grant numbers 12525414 and 62235006 to X.-W.C., 12374349 to J.T., 52272270 and 51972084 to G.C.), the Fundamental Research Funds for the Central Universities (grant numbers 2023BR003 to X.-W.C., 2024BRB002 to J.T., AUGA5710052614 and AUGA8880100415 to G.C.), the Hubei Provincial Natural Science Foundation of China (grant number 2026AFA092 to J.T.), the Hubei Provincial Talent Program (J.T.), and the Fundamental Research Fund for Distinguished Scholars (Harbin Institute of Technology) (grant number XWQQ5710001615 to G.C.).

**Author contributions**

X.-W.C. initiated the research. X.-W.C., G.C., J.T., Y.M. and T.M. designed the experiments. T.M. and Y.M. carried out the optical measurements and super-resolution localization microscopy experiment. D.H. and F. L. synthesized and characterized the UCNC samples under the guidance of G.C.. X.-W.C., J.T., and T.M. discussed the results of optical measurements and analyzed data. T.M., J.T. and X.-W.C. wrote the paper with inputs from all other authors. The project was supervised by X.-W.C., J.T. and G.C.

**Competing interests**

The authors declare no competing interests.

**Additional information**

Correspondence and requests for materials should be addressed to X.-W.C., J.T. or G.C.

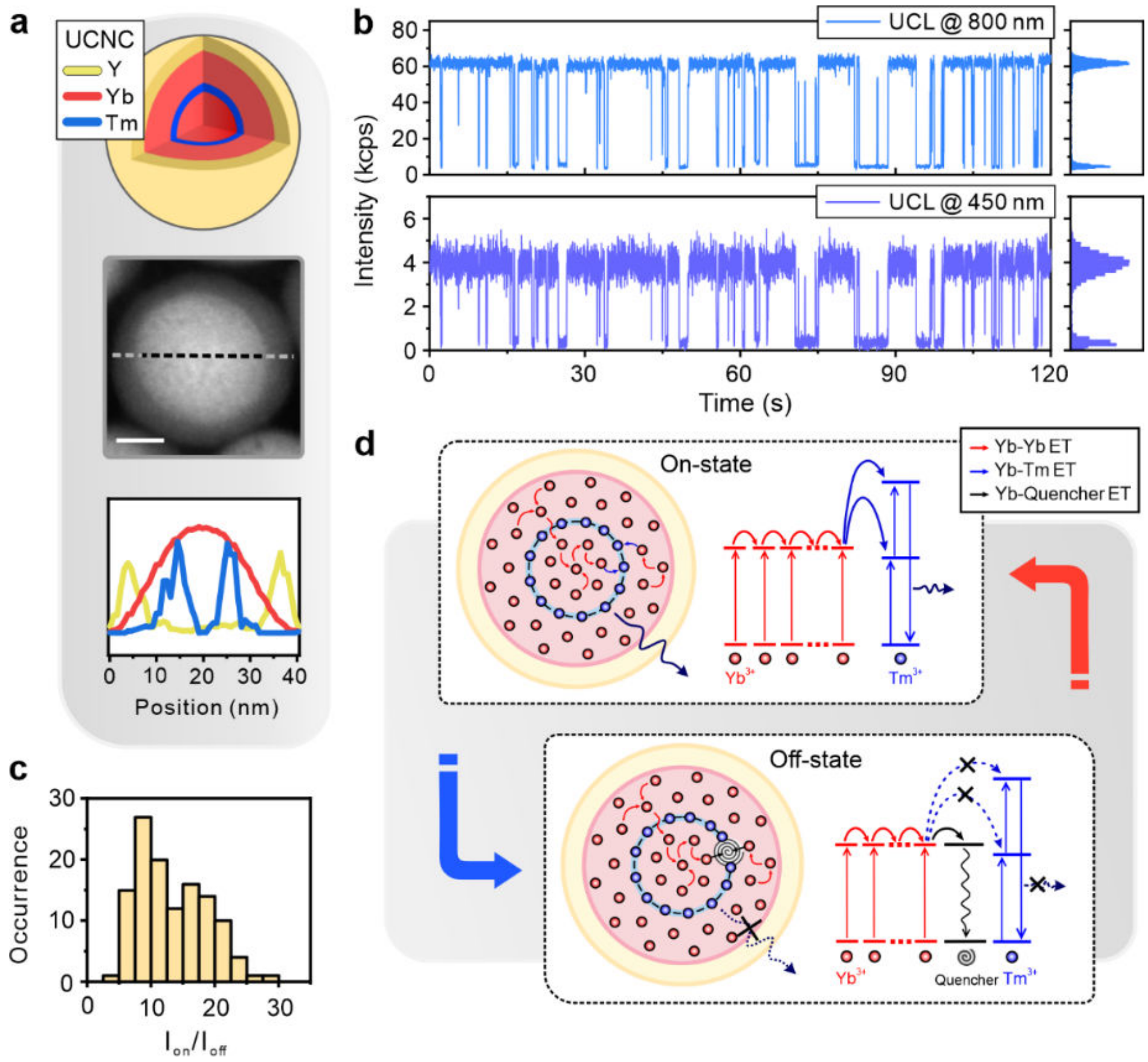

**Figure 1 | UCL blinking and the proposed blinking mechanism. a**, Top: structural model of the core@shell@shell@shell ($NaYbF_4$@$NaTmF_4$@$NaYbF_4$@$NaYF_4$) UCNC. Middle: dark-field scanning transmission electron microscopy (TEM) image. Scale bar, 10 nm. Bottom: Energy dispersive spectrometry (EDS) profile along the dashed line marked in the TEM image. **b**, Intensity time traces of the long-wavelength (~ 800 nm) and short-wavelength (~ 450 nm) UCL bands from a single UCNC, exhibiting synchronous blinking. Intensity distribution histograms are plotted to the right of the traces. The excitation power density is 2.3 kW/cm$^2$. **c**, Histogram of the on-off intensity ratio (mean value 13.6) obtained from 122 individual UCNCs. **d**, Schematic illustration of the proposed mechanism for UCL blinking. See text for details.

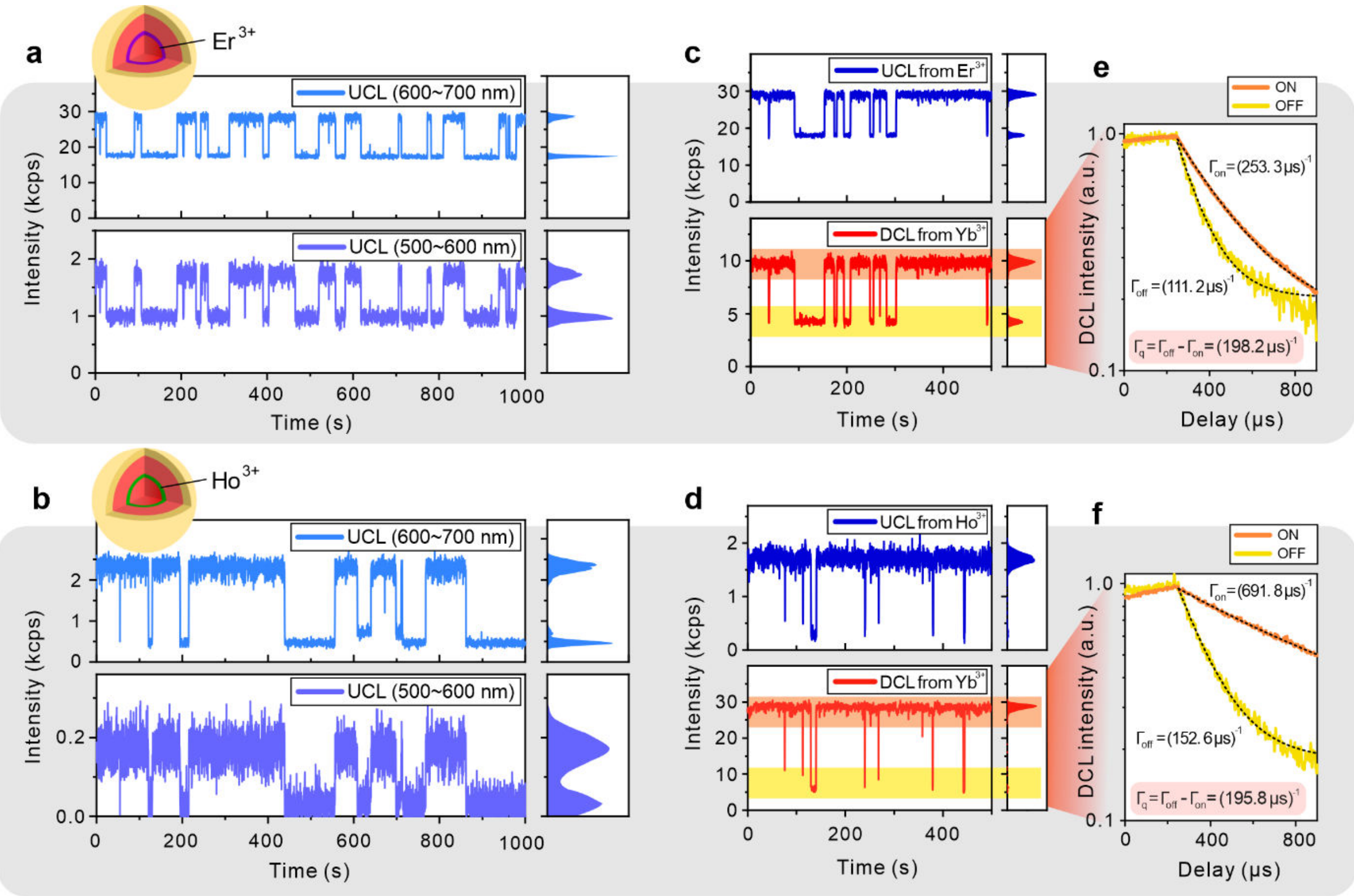

**Figure 2 | Synchronous UCL and DCL blinking in systems doped with $Er^{3+}$ and $Ho^{3+}$. a, b**, Intensity time traces of the long-wavelength (600~700 nm) and short-wavelength (500~600 nm) UCL bands from single $Er^{3+}$-doped (**a**) and $Ho^{3+}$-doped (**b**) UCNCs, exhibiting synchronous blinking. **c**, **d**, Intensity time traces of the UCL and DCL from single $Er^{3+}$-doped (**c**) and $Ho^{3+}$-doped (**d**) UCNCs, revealing synchronous blinking of the UCL and DCL. Corresponding intensity distribution histograms are shown to the right of each trace. **e**, **f**, Decay dynamics of the on-state (orange curve) and off-state (yellow curve) DCL from $Yb^{3+}$ in $Er^{3+}$-doped UCNC (**e**) and $Ho^{3+}$-doped UCNC (**f**). The extracted decay rates $\Gamma_{\mathrm{on}}$ and $\Gamma_{\mathrm{off}}$ are shown alongside the decay curves. The quenching rates $\Gamma_{\mathrm{q}}$ (calculated as $\Gamma_{\mathrm{q}} = \Gamma_{\mathrm{off}} - \Gamma_{\mathrm{on}}$) are also provided and exhibit excellent consistency between the two systems.

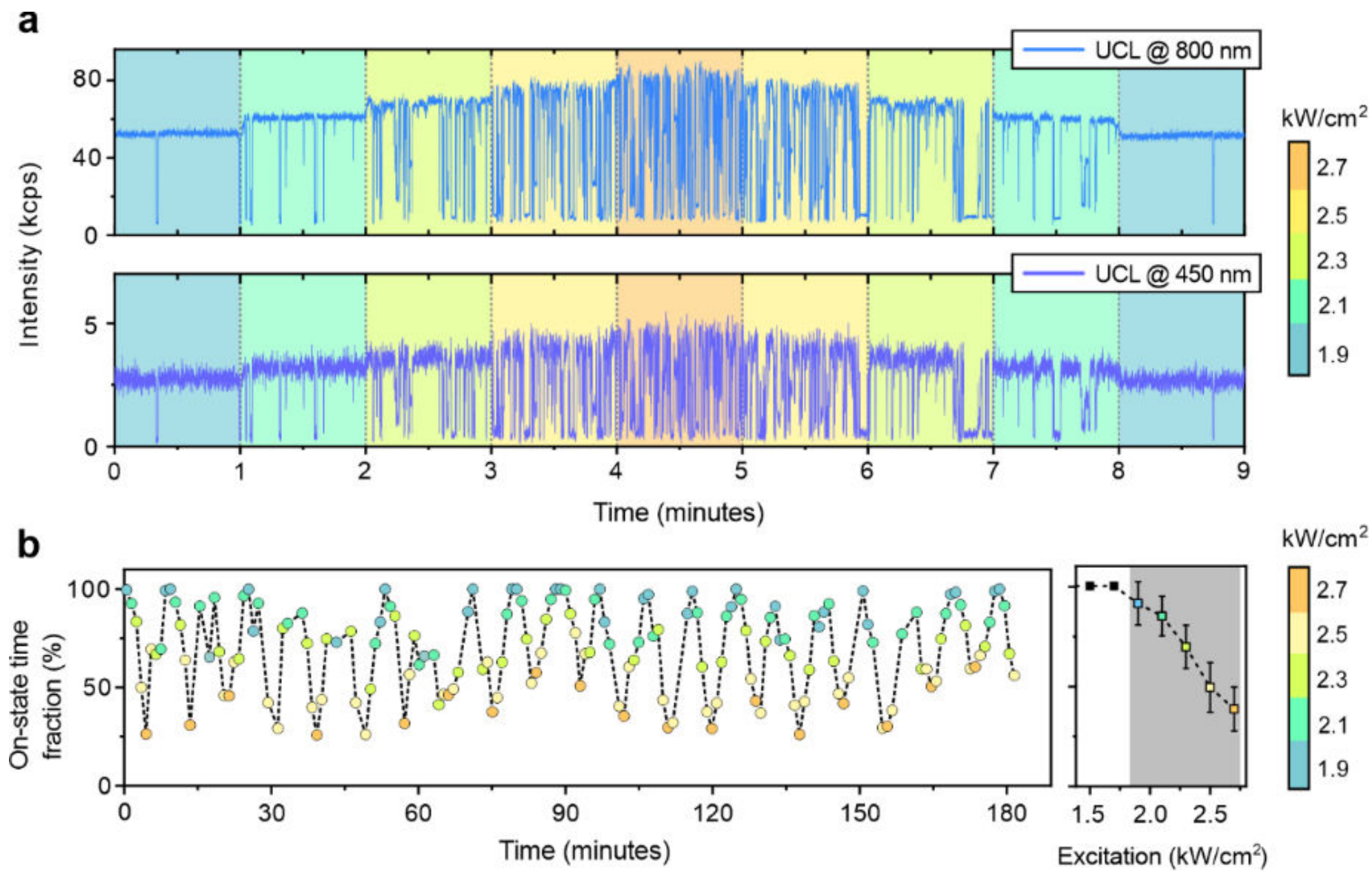

**Figure 3 | Blinking control via the excitation power. a**, Evolution of the UCL blinking behavior of a single UCNC during an excitation power cycle, in which the excitation power density is gradually increased from 1.9 kW/cm$^2$ to 2.7 kW/cm$^2$ and then decreased back to 1.9 kW/cm$^2$ (in steps of 0.2 kW/cm$^2$). Background colors indicate the excitation power densities as defined in the color bar. **b**, On-state fraction as the excitation power density cycles between 1.9 kW/cm$^2$ and 2.7 kW/cm$^2$ for over 3 hours. The average on-state fraction as a function of excitation power density is plotted on the right side, with error bars representing one standard deviation. The colors of the data points indicate the excitation power densities as defined in the color bar.

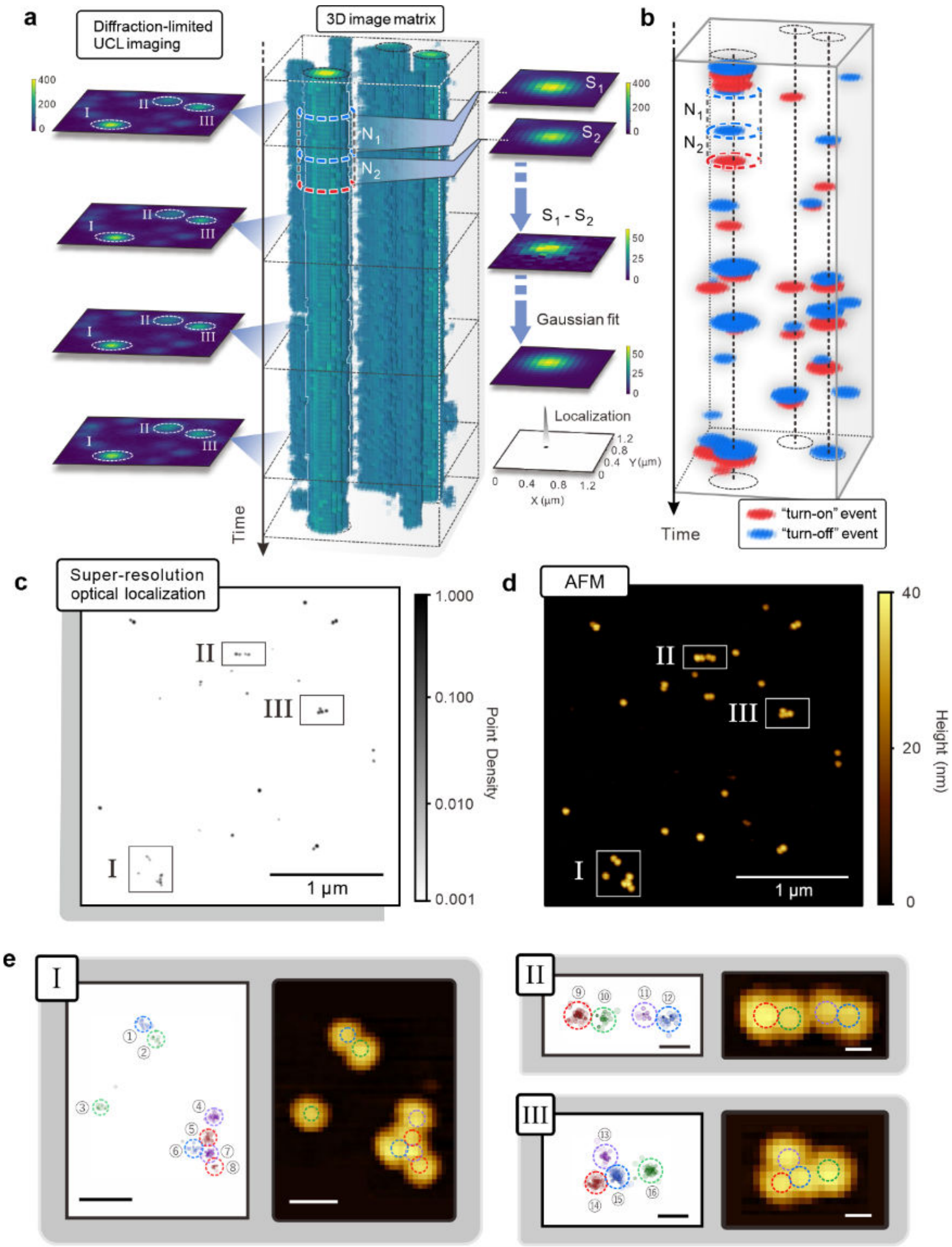

**Figure 4 | Super-resolution localization microscopy based on UCL blinking. a**, A time series of 3×3 μm$^2$ wide-field UCL image frames arranged into a 3D (*x* , *y*, *t*) image matrix. Each frame is acquired with an integration time of 250 milliseconds. Four representative frames are displayed in the left column. Three brightest luminescence spots containing multiple luminescent UCNCs are marked as "I", "II", and "III". The right column illustrates the procedure to resolve and localize a single UCNC from a dense UCNC aggregate. **b**, Identification of the blinking events in the time-differential image matrix obtained via subtracting each UCL image frame with the preceding one. Clusters of positive (red) and negative (blue) matrix elements correspond to the "turn-on" and "turn-off" blinking events of individual UCNCs within a luminescence spot. Columnar regions $N_1$ and $N_2$ represent two consecutive non-blinking intervals between temporally adjacent blinking events. **c**, Super-resolution optical localization image, obtained by applying Gaussian rendering on the 2D localization point cloud. The spatial region is the same as the wide-field UCL images shown in panel **a**. **d**, AFM topographic image for the same region as panel **c**. **e**, Magnified optical localization images and the corresponding AFM topographic images for the three densest regions. The optical localization points are shown as circles (size indicates 1σ Gaussian fitting error) and assigned to individual UCNCs via clustering analysis using Gaussian mixture model. Dashed circles are centered at the weighted centroids of these clusters to represent particle locations, with their diameters corresponding to the particle heights measured via AFM. Scalar bars for region "I", 100 nm; Scalar bars for regions "II" and "III", 40 nm.

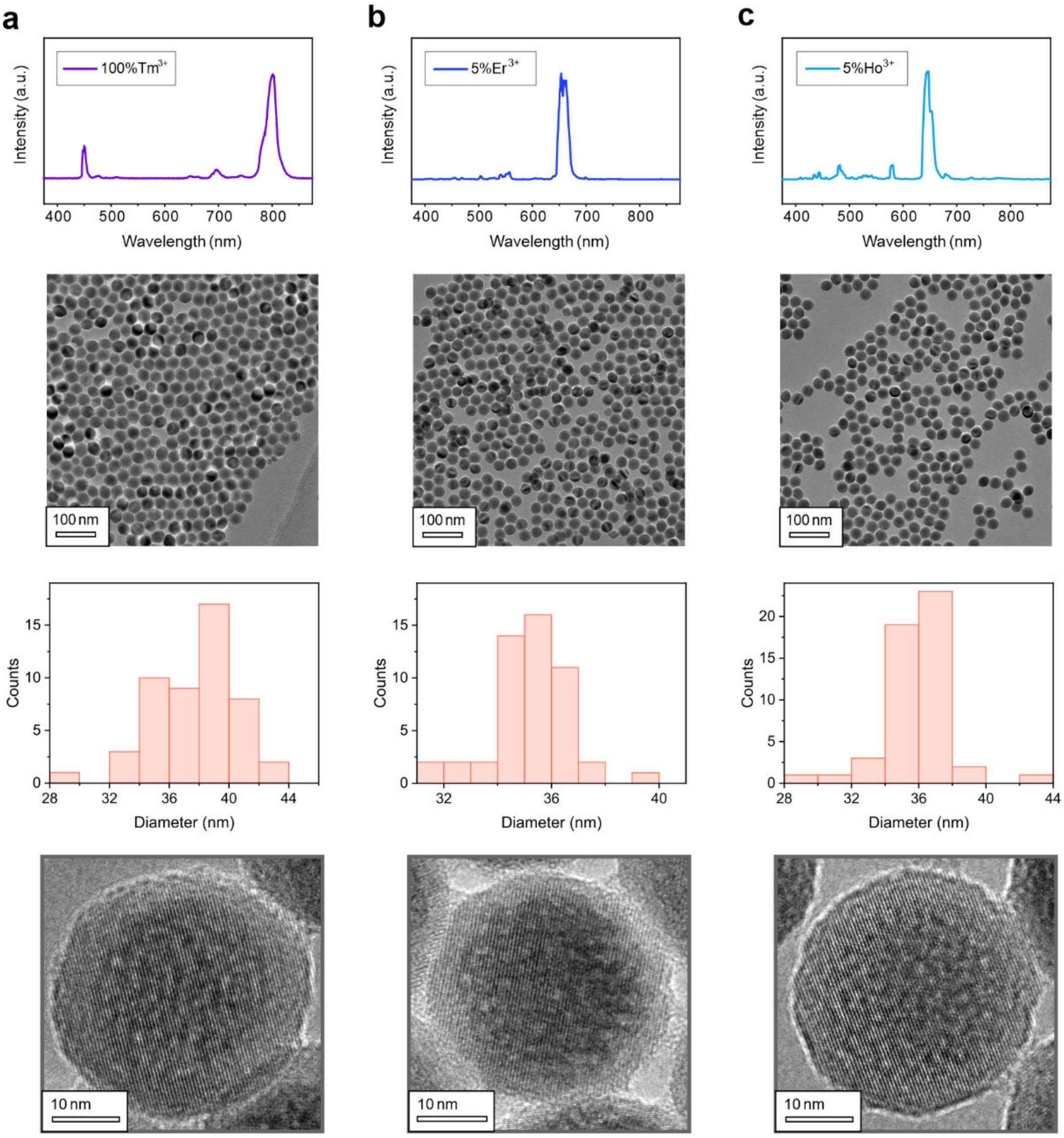

**Extended Data Fig. 1 | Spectroscopic and structural characterization of the UCNC samples. a-c** Spectroscopic and structural characterization of the $Tm^{3+}$-doped (**a**), $Er^{3+}$-doped (**b**) and $Ho^{3+}$-doped (**c**) UCNCs. 1st row: Upconversion luminescence (UCL) spectrum. 2nd row: Transmission electron microscopy (TEM) image. 3rd row: Size-distribution derived from the TEM image. 4th row: High-resolution transmission electron microscopy (HR-TEM) image.

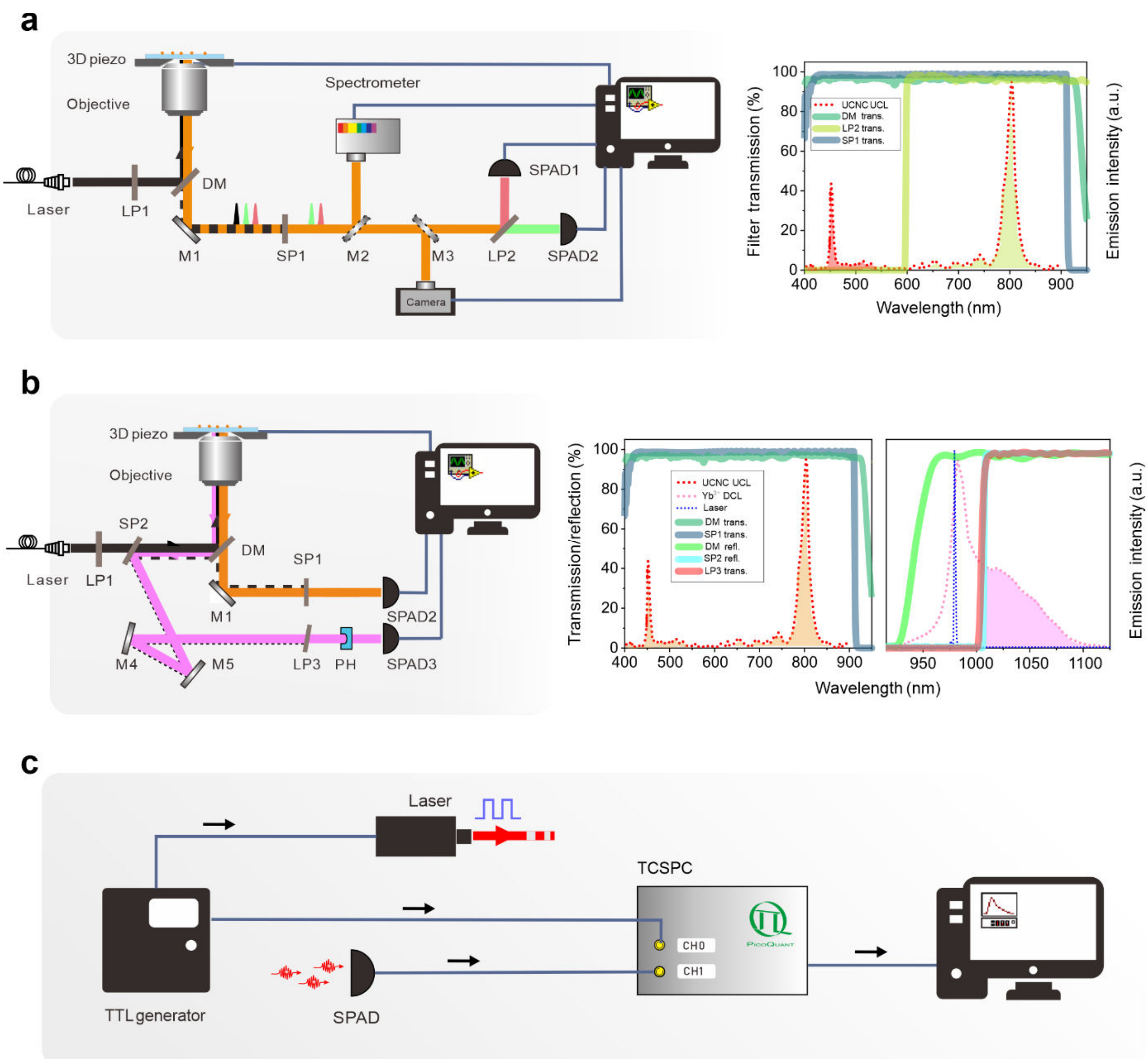

**Extended Data Fig. 2 | Experimental setups. a**, Setup (left) and spectral filtering configuration (right) for UCL measurement. DM: dichroic mirror; M1: mirror; M2, M3: flip mirrors; SP1: short-pass filter; LP1, LP2: long-pass filters; SPAD1, SPAD2: single-photon avalanche photodiodes. Black line: excitation light. Dashed black line: excitation light leakage. Orange line: UCL. Red line: long-wavelength component of the UCL. Green line: short-wavelength component of the UCL. The right panel shows the transmission spectra of the filters employed in the optical setup, along with the UCL spectrum, illustrating the spectral filtering configuration for UCL measurement; The red and green shaded areas of the UCL spectrum indicate the long-wavelength and short-wavelength components of the UCL, respectively. **b**, Setup (left) and spectral filtering configuration (right) for correlative measurement of UCL and DCL. SP2: short-pass filter; LP3: long-pass filter; PH: pinhole. Orange line: UCL light. Pink line: DCL light. The right panel shows the transmission/reflection spectra of the filters employed in the optical setup, along with the UCL and DCL spectrum, illustrating the spectral filtering configuration for correlative detection of UCL and DCL; the orange and pink shaded areas of the spectrum indicate the UCL and DCL spectra, respectively. **c**, Setup for fluorescence lifetime measurement. A TTL generator synchronizes the pulse modulation of the 980 nm laser with the TCSPC module. Emitted photons are detected by the SPAD. The time interval between the laser trigger (“CH0”) and the photon detection event (“CH1”) is registered to reconstruct the time-resolved fluorescence decay histogram.

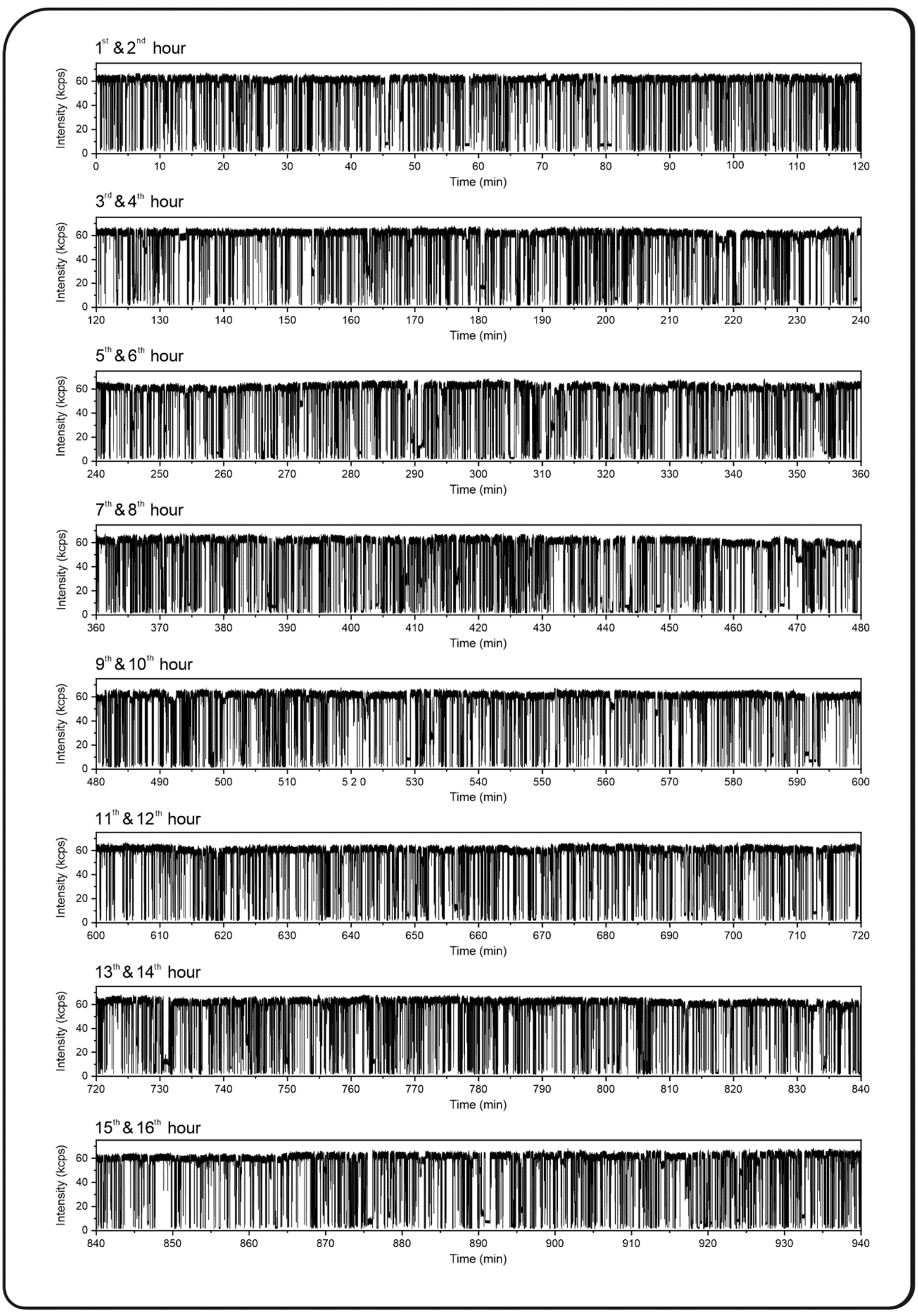

**Extended Data Fig. 3 | UCL time trace of a single UCNC for over 15 hours.** The time trace was recorded under an excitation power density of 2.2 kW/cm$^2$. Manual realignment was performed every 30 minutes during the measurement to compensate for sample drift. There are more than 10,000 blinking cycles throughout the observation period of 940 minutes.

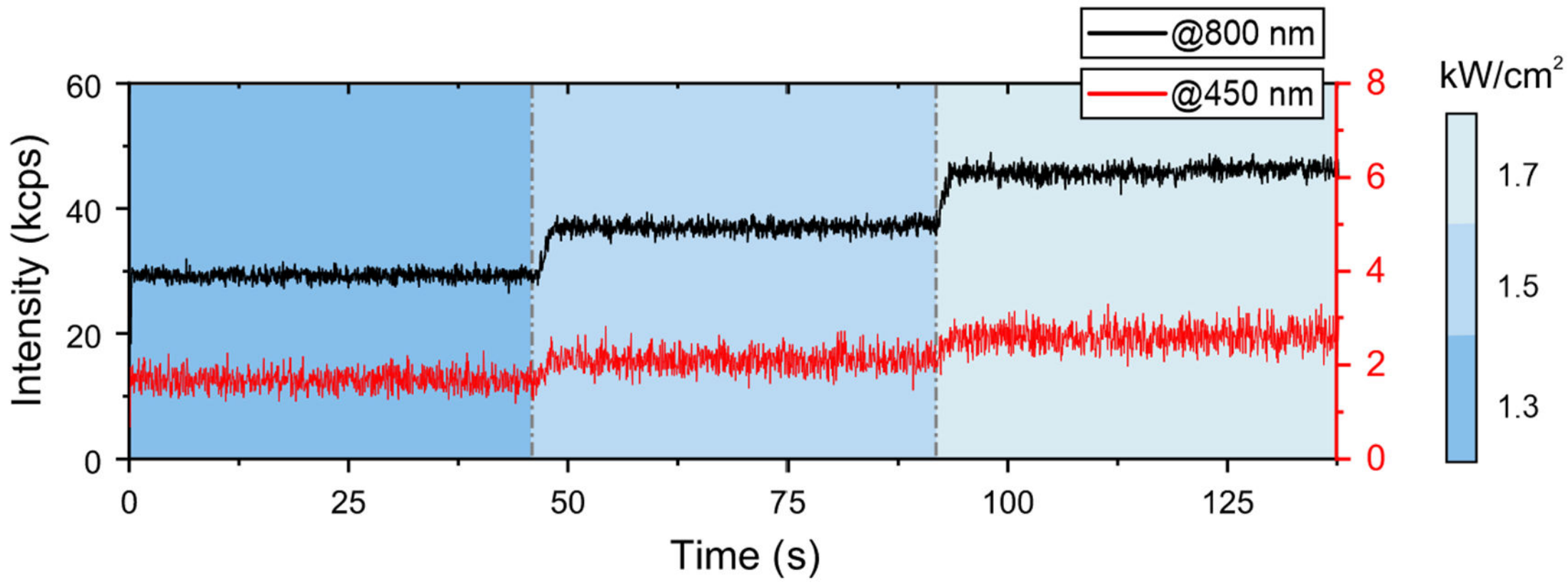

**Extended Data Fig. 4 | Non-blinking UCL time traces of a single UCNC at the excitation power densities below the threshold for blinking.** The UCL time traces from left to right were recorded at the excitation power densities of 1.3 kW/cm$^2$, 1.5 kW/cm$^2$, and 1.7 kW/cm$^2$, respectively, as indicated by different background colors. All these excitation power densities are below the threshold for blinking. The black and red traces represent the UCL spectral components around 800 nm and 450 nm, respectively.

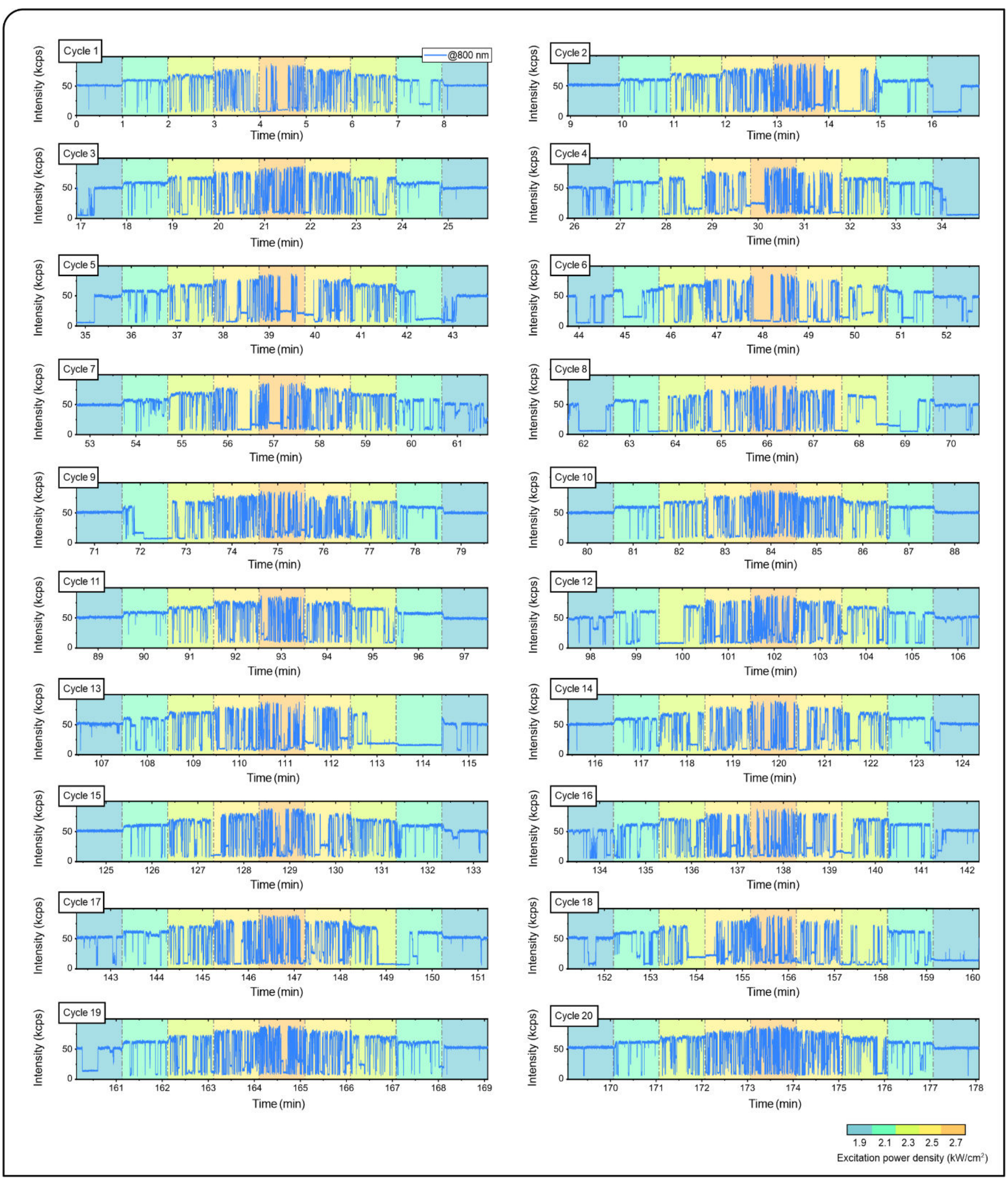

**Extended Data Fig. 5 | UCL blinking time trace of a single UCNC across 20 cycles of periodic modulation of the excitation power.** Continuous 3-hour UCL time trace of a single UCNC recorded as the excitation power density repeatedly varied between 1.9 and 2.7 kW/cm$^2$. Background shading indicates the excitation power levels defined in the color bar. From this time trace, we extract the evolution of the on-state time fraction, which is displayed in Fig. 3b of the main text. In the calculation of on-state time fractions, off-states longer than 15 seconds have been excluded, as the limited 1-minute integration time per data point means such occasional prolonged dark intervals would introduce substantial deviations.

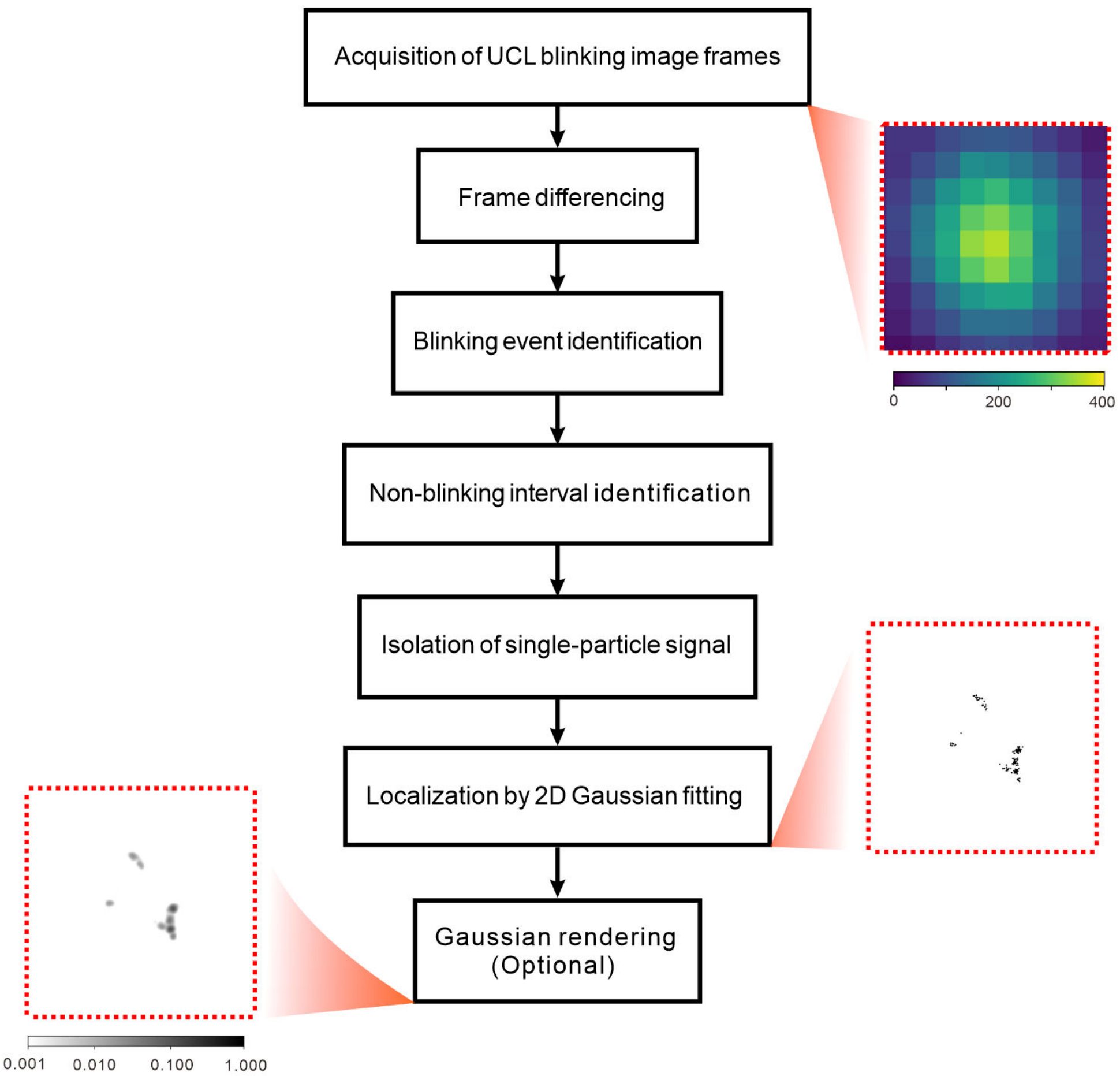

**Extended Data Fig. 6 | Flowchart of the super-resolution localization algorithm.** Schematic workflow for reconstructing super-resolved images from raw UCL image frames. The process initiates with raw data acquisition and frame differencing. Sequential steps identify stochastic blinking events and non-blinking intervals, enabling isolation of high-fidelity single-particle emission spots. Spatial coordinates of the single particles are then determined via 2D Gaussian fitting of the single-particle emission spots. The localizations are corrected for sample drift and compiled into a 2D point cloud. As an optional visualization step, the localization point cloud can be converted into a continuous intensity map via Gaussian rendering. The insets show representative results from this super-resolution algorithm applied to the UCNC cluster corresponding to spot "I" in Fig. 4.

**Extended Data Table 1 | Localization precision for each single UCNC.**

| UCNC # (numbered in Fig. 4e) | Number of localization points | $\sigma_X$ (nm) | $\sigma_Y$ (nm) |
|---|---|---|---|
| **1** | 16 | 1.9780 | 1.3172 |
| **2** | 9 | 1.7039 | 2.8274 |
| **3** | 9 | 2.0636 | 1.0059 |
| **4** | 36 | 1.0547 | 0.8031 |
| **5** | 28 | 0.9113 | 1.2521 |
| **6** | 13 | 1.7226 | 1.3504 |
| **7** | 32 | 0.7476 | 0.6536 |
| **8** | 8 | 0.9937 | 1.0187 |
| **9** | 36 | 0.9001 | 0.8231 |
| **10** | 22 | 1.3134 | 1.0839 |
| **11** | 18 | 1.4350 | 3.0535 |
| **12** | 8 | 1.4674 | 1.4728 |
| **13** | 27 | 0.6633 | 0.9117 |
| **14** | 50 | 0.7870 | 0.6759 |
| **15** | 69 | 0.5654 | 0.6341 |
| **16** | 46 | 0.8246 | 0.6060 |
| Mean value | | 1.1957 | 1.2181 |